\documentclass{article}

\usepackage{1_arxiv}     

\usepackage[utf8]{inputenc} 
\usepackage[T1]{fontenc}    
\usepackage{url}            
\usepackage{booktabs}       
\usepackage{amsfonts}       
\usepackage{nicefrac}       
\usepackage{microtype}      
\usepackage{lipsum}
\usepackage{float}
\usepackage{natbib}         
\usepackage{amsmath}        
\usepackage{fancyhdr}       
\usepackage{graphicx}       
\usepackage{siunitx}        
\usepackage{multirow}       
\usepackage{hyperref}       

\graphicspath{{img/}}       

\title{
    Hybrid Optical Turbulence Models Using Machine Learning and Local Measurements\thanks{Cite as: Applied Optics 62(18) 4880-4890, doi: 10.1364/AO.487280}
}
\date{April 27, 2023}

\author{ \href{https://orcid.org/0000-0003-0469-353X}{\includegraphics[scale=0.06]{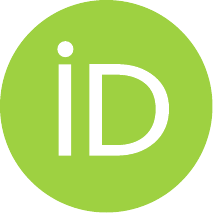}\hspace{1mm}Christopher ~Jellen}\thanks{Corresponding Author} \\
    Mechanical Engineering Department\\
	Untied States Naval Academy\\
	Annapolis, MD 21402 \\
	\texttt{cdjellen@gmail.com} \\
	\And
	Charles ~Nelson \\
	Electrical Engineering Department\\
	Untied States Naval Academy\\
	Annapolis, MD 21402 \\
	\And
	John ~Burkhardt \\
    Mechanical Engineering Department\\
	Untied States Naval Academy\\
	Annapolis, MD 21402 \\
    \And
	Cody ~Brownell \\
    Mechanical Engineering Department\\
	Untied States Naval Academy\\
	Annapolis, MD 21402 \\
}

\renewcommand{\shorttitle}{Hybrid Models for Optical Turbulence}

\hypersetup{
pdftitle={Hybrid Optical Turbulence Models Using Machine Learning and Local Measurements},
pdfsubject={physics.ao-ph, cs.ai},
pdfauthor={Christopher ~Jellen, Charles ~Nelson, John ~Burkhardt, Cody ~Brownell},
pdfkeywords={hybrid model, optical turbulence, boundary layer, scintillation, machine learning},
}

\begin{document} 
\maketitle
\setcitestyle{numbers}
\begin{abstract} 

Accurate prediction of atmospheric optical turbulence in localized environments is essential for estimating the performance of free-space optical systems. Macro-meteorological models developed to predict turbulent effects in one environment may fail when applied in new environments. However, existing macro-meteorological models are expected to offer some predictive power. Building a new model from locally-measured macro-meteorology and scintillometer readings can require significant time and resources, as well as a large number of observations.  These challenges motivate the development of a machine-learning informed  hybrid model framework. By combining some baseline macro-meteorological model with local observations, hybrid models were trained to improve upon the predictive power of each baseline model. Comparisons between the performance of the hybrid models, the selected baseline macro-meteorological models, and machine-learning models trained only on local observations highlight potential use cases for the hybrid model framework when local data is expensive to collect. Both the hybrid and data-only models were trained using the Gradient Boosted Decision Tree (GBDT) architecture with a variable number of in-situ meteorological observations. The hybrid and data-only models were found to outperform three baseline macro-meteorological models, even for low numbers of observations, in some cases as little as one day. For the first baseline macro-meteorological model investigated, the hybrid model achieves an estimated \si{29\%} reduction in mean absolute error (MAE) using only one days-equivalent of observation, growing to \si{41\%} after only two days, and \si{68\%} after \si{180} days-equivalent training data. The data-only model generally showed similar but slightly lower performance as compared to the hybrid model. Notably, the hybrid model’s performance advantage over the data-only model dropped below \si{2\%} near the 24 days-equivalent observation mark and trended towards \si{0\%} thereafter. The number of days-equivalent training data required by both the hybrid model and the data-only model is potentially indicative of the seasonal variation in the local microclimate and its propagation environment.  

\end{abstract}

\section{Introduction} 

Atmospheric optical turbulence degrades the performance of free-space optics (FSO) and other optical systems, especially at low altitudes and in the near-maritime environment \cite{barrios2012wireless} \cite{wang2015prediction} \cite{frederickson2000estimating} \cite{frederickson2006measurements}. These effects are characterized by the refractive index structure parameter, $C_n^2$. For horizontal propagation, under the assumption of isotropy and path-wise homogeneity, fluctuations in $C_n^2$ are dominated by temperature fluctuations \cite{barrios2012wireless}. The impact of atmospheric factors on $C_n^2$ led to the development of models which predict local turbulent effects from macro-meteorological features \cite{wang2015prediction} \cite{sadot1992forecasting} \cite{raj2015comparison} \cite{chen2019climatological}. 

Existing macro-meteorological models are often extended to new microclimates in an attempt to generate optical turbulence predictions using local atmospheric feature measurements. These models may generate predictions with higher error when applied to these new microclimates than in the environment in which the model was originally developed \cite{wang2015prediction} \cite{jellen2020machine} \cite{jellen2020measurement} \cite{oermann2014novel}. Some state-of-the-art models have performed well across similar microclimates, including NAVSLaM which performed well for both coastal and near-maritime propagation paths \cite{mahon2020comparison}. However, the equipment required to effectively measure potential temperature and wind shear gradients for state-of-the-art model predictions are often unavailable or cost prohibitive \cite{wang2015prediction}. Additionally, developing a new model for each microclimate can often be more costly in time, equipment, and expertise, when existing models may hold some predictive power across a range of propagation environments, and may have the potential for augmentation rather than full redevelopment. 

These challenges motivate investigation into the development of the hybrid model framework.  The hybrid model couples existing macro-meteorological models developed for similar microclimates along with some minimal amount of locally-acquired meteorological and $C_n^2$ data. The hybrid model framework consists of two components, a baseline macro-meteorological model and a machine learning model trained on that baseline macro-meteorological model’s residual error over the locally-acquired training measurements. Under this approach, the hybrid model’s predictions serve as the baseline macro-meteorological model’s predictions and augmented by a correction learned from locally-measured meteorological and $C_n^2$ data. In this paper the hybrid model approach is investigated with great detail for one specific baseline macro-meteorological model as well as for a single propagation path; however, the hybrid model approach itself is not presented in this paper as being specific to any one model, architecture, or microclimate.  To that end, two additional baseline models are evaluated to demonstrate the extensibility and potential limitations of hybrid model approach. 

Using locally-acquired scintillometer and weather station measurements collected over the Severn River for approximately 31 months, a hybrid model was developed and compared against both a data-only model trained under the Gradient Boosting Decision Tree (GBDT) architecture in \cite{ke2017lightgbm} using local measurements, and one specific baseline macro-meteorological model developed for a similar microclimate and presented in \cite{chen2019climatological}. This particular macro-meteorological model was chosen as a baseline model as comparison as it was developed for a propagation path both over water, and also had a similar emphasis on the air-water temperature difference along the path, as was also done in \cite{jellen2020machine}. Both the hybrid and data-only models were trained with a variable number of bootstrapped training observations to investigate the marginal improvement in prediction accuracy resulting from more training observations.  

The hybrid model framework demonstrated a significant improvement in predictive performance when compared to the baseline macro-meteorological models in \cite{wang2015prediction} \cite{sadot1992forecasting} \cite{chen2019climatological}; even with a very limited number of training observations. For example, in comparison with the baseline model presented in \cite{chen2019climatological}, and with only one days-equivalent data observation, the hybrid model demonstrated an estimated \si{29\%} improvement in mean absolute error (MAE) in predicting $\log_{10} C_n^2$. Additionally, the hybrid model’s performance improved steadily, with an estimated \si{68\%} improvement using 180 days-equivalent of observation, and then saw marginal improvement thereafter.

\section{Method}

The principle aim of this study was to develop and analyze the hybrid model framework, which combines a baseline macro-meteorological model for predicting $C_n^2$ from local measurements with a machine learning model trained to predict the baseline model’s residual error in the local microclimate. For this study, the GBDT architecture was selected to learn the baseline model’s residual error and the hybrid model was compared to a data-only model of similar architecture trained using only local measurements.

The GBDT component of the hybrid model was trained on training observations of meteorological parameters and a target correction, $tc$, tailored to the baseline model’s predicted $\log_{10} C_n^2$, as expressed in equation (\ref{eq1}).

\begin{equation}\label{eq1}
    tc=\log{10}C^2_{n_{observed}}-\log{10}C^2_{n_{baseline\ model\ predicted}}
\end{equation}

The GBDT learns a functional approximation for $tc$, and for observations outside of the training set, generates a predicted \(\hat{tc}\). The baseline model’s predicted $\log_{10} C_n^2$ is then adjusted using equation (\ref{eq2}). 

\begin{equation}\label{eq2}
    \log{10}C^2_{n_{hybrid\ model\ predicted}}=\log{10}C^2_{n_{baseline\ model\ predicted}}+tc
\end{equation}

The predicted $\log_{10} C_n^2$ generated by the hybrid model is thus a composite of the baseline macro-meteorological model’s prediction and the GBDT component’s predicted \(\hat{tc}\). This prediction is the hybrid model’s output given an observed vector of macro-meteorological measurements. Under this framework, the hybrid model’s GBDT component seeks to learn the mapping between local meteorological data and the baseline model’s residuals, such that the composite prediction in equation (\ref{eq2}) demonstrates lower error than the baseline model alone. 

A secondary aim of this study was to determine the amount of locally-acquired data required to train an effective hybrid model or a data-only model under the GBDT architecture. While the macro-meteorological parameters used to predict local $C_n^2$ are readily-available, measured both by existing weather stations \cite{noaandbctplm22022} or by commercial off the shelf hardware \cite{davisvantagepro22023}, the hybrid model framework does require some number of locally-acquired $C_n^2$ measurements to learn appropriate local corrections. Determining the minimum number of required observation days to achieve some estimated improvement in performance, as well as the relationship between long-term prediction error and number of training observations, will aid in operationalizing the hybrid model approach to new contexts and microclimates. 

\section{Measurement} 
\subsection{Data Collection} 

To investigate optical turbulence in the low-altitude near-maritime environment, an \si{890\metre} propagation path was established over the Severn River in Annapolis, Maryland. A scintillometer was used to establish a measure of optical turbulence, $C_n^2$, with which to compare model predictions. The scintillometer link was approximately horizontal, with an elevation of \si{2\metre} to \si{4\metre} over the surface of the water depending on tides. The average elevation of the link was estimated at \si{3\metre}. Significant landmasses exist at either end of the propagation path, however, approximately \si{98\%} of the path is over water. This environment has been previously characterized as “near-maritime” and “littoral”, distinct from open-ocean propagation environments and paths exclusively over land \cite{frederickson2000estimating} \cite{frederickson2006measurements} \cite{jellen2020machine} \cite{jellen2020measurement} \cite{Jellen:21}.

\begin{figure}[H]
    \centering
    \includegraphics[width=0.9\textwidth]{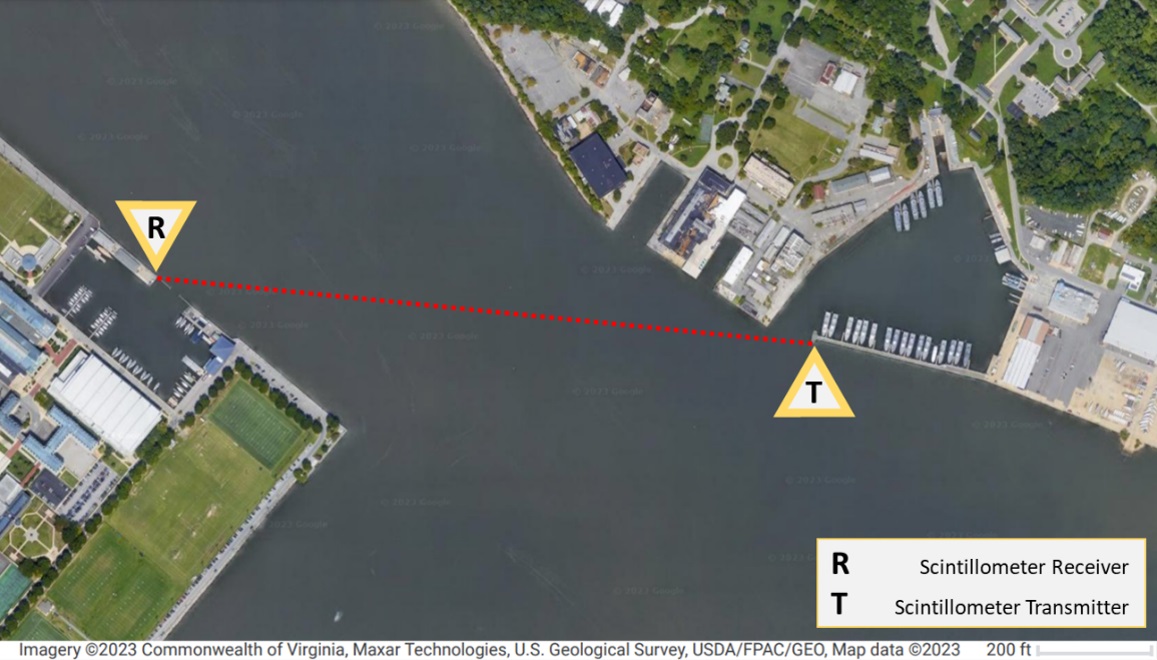}
    \caption{Scintillometer propagation path with receiver (R) and transmitter (T) for the BLS 450 scintillometer across the Severn River in Annapolis, Maryland \cite{googlemapsannapolis2023}.}
    \label{fig:1}
\end{figure}

$C_n^2$ measurements and associated timestamps were captured across the BLS 450 scintillometer link pictured in \autoref{fig:1}. This link provided measures of optical turbulence for approximately 31 months and were used to develop the hybrid model, the data-only model, and to make comparisons with the selected baseline macro-meteorological model. A local weather station was deployed next to the receiver which captured macro-meteorological parameters such as air temperature, wind speed, pressure, humidity, and solar radiation \cite{davisvantagepro22023}. Additionally, publicly available data from the nearest NDBC data buoy was used to obtain hourly-averaged water temperature readings for the local environment \cite{noaandbctplm22022}. More information about each of these data sources and their methodologies is available in \cite{jellen2020machine} \cite{jellen2020measurement} \cite{noaandbctplm22022}. The elevation of the local weather station was approximately \si{3\metre} above the mean lower low water line, with water temperature readings captured approximately \si{1\metre} below the mean lower low water line \cite{noaandbctplm22022}. The measurements captured are described in \autoref{table:1}. 
 
\begin{table}[H]
    \centering
    \caption{
        Macro-meteorological and oceanographic parameters captured near the link (January 1\textsuperscript{st} 2020 through July 14\textsuperscript{th} 2022).
    }
    \begin{tabular}{|| m{4cm} || m{2.8cm} | >{\centering}m{1.2cm} | >{\centering}m{2cm} | >{\centering\arraybackslash}m{2.5cm} ||}
        \hline
        \textbf{Parameter} & \textbf{Data Source} & \textbf{Unit} & \textbf{Measurement Frequency} & \textbf{Number of Observations} \\
        \hline
        \hline
        \begin{tabular}
            [c]{@{}l@{}}$C_n^2$\end{tabular} & BLS 450 & \si{\metre\textsuperscript{$-\frac{2}{3}$}} & \si{1\minute} & \si{1,246,802} \\
            \hline
            Water Temperature & NDBC Data Buoy & \si{\celsius} & \si{1\hour} & \si{21,946} \\
            \hline
            Air Temperature & \multirow{5}{*}{Vantage Pro2} & \si{\celsius} & \multirow{5}{*}{\si{10\minute}} & \si{125,569} \\
            Atmospheric Pressure &  & \begin{tabular}[c]{@{}l@{}} \si{\milli\bar} \end{tabular} &  & \si{126,510} \\
            Relative Humidity &  & \begin{tabular}[c]{@{}l@{}} \si{\%} \end{tabular} &  & \si{125583} \\
            Solar Radiation &  & \si{\watt\per\squared\metre} &  & \si{125,633} \\
            Wind Speed &  & \begin{tabular}[c]{@{}l@{}} \si{\metre\per\second} \end{tabular} &  & \si{126,512} \\
        \hline                                                                                          
    \end{tabular}
    \label{table:1}
\end{table}

Before evaluating macro-meteorological models and developing hybrid models for the local propagation environment, the macro-meteorological and oceanographic parameters in \autoref{table:1} were re-sampled and interpolated to provide \si{1\minute} estimates of each parameter of interest. Additionally, the temporal hour for each observation was calculated by subtracting the time captured by the BLS 450 scintillometer from that day’s sunrise time, and the air-water temperature difference in \si{\celsius} was computed for each applicable observation.  
 
\section{Model training and evaluation}

\subsection{Baseline macro-meteorological model}

The literature is rich with examples of regression-based models for predicting local optical turbulence effects from meteorological parameters \cite{wang2015prediction} \cite{sadot1992forecasting} \cite{raj2015comparison} \cite{chen2019climatological}. These studies often seek to develop models based on macro-meteorology for their local propagation environment. Additionally, macro-meteorological models have shown some promise in generating predicted $C_n^2$ when applied to new propagation environments \cite{wang2015prediction}. While these models tend to demonstrate impressive predictive power when employed in the environment in which they were developed, performance tends to degrade when models are applied in other propagation environments or over longer periods of time \cite{wang2015prediction} \cite{Jellen:21}. 

The near-maritime propagation environment at the United States Naval Academy has some seasonal variation in measured $C_n^2$, with a predicted dependence on the temperature difference between the air and water in the air-to-water boundary layer above the Severn River \cite{frederickson2006measurements} \cite{jellen2020machine}. While this propagation path has an elevation near sea level, these characteristics are similar to other propagation environments over water, where seasonal variation in $C_n^2$ has been observed. The boundary-layer propagation path over Fuxian Lake, as measured at the Fuxian Solar Observatory, is one such environment \cite{chen2019climatological}. Both locations have land masses at each end of the propagation path and are relatively flat in the immediate vicinity of the scintillometer link.  

Macro-meteorological models were developed to predict boundary layer $C_n^2$ from local measurements at the Fuxian Solar Observatory \cite{chen2019climatological}. The authors of \cite{chen2019climatological} reference a model for predicting ground-level $C_n^2$ for “normal meteorological conditions” at Fuxian Solar Observatory as reproduced in equation (\ref{eq3}) \cite{chen2019climatological} 

\begin{equation}\label{eq3}
    C_n^2(0) = ({2.05\Delta T}^2 + 2.37\Delta T + 1.58) \times 10^{-16}
\end{equation}

In equation (\ref{eq3}), the predicted $C_n^2(0)$ is a function of only the measured air-water temperature difference \(\delta T\). The value for equation (\ref{eq3}) at a height $h$ in \si{\metre} within the boundary layer can be calculated using equation (\ref{eq4}) \cite{chen2019climatological} \cite{fried1966optical} 

\begin{equation}\label{eq4}
    C_n^2(h) = C_n^2(0)h^{-1/3}e^{\frac{-h}{h_0}}
\end{equation}

In equation (\ref{eq4}), $h_0$ is a constant equal to \si{3200\metre} where $h$ is a height within the local boundary layer. For the propagation path over the Severn River, $h$ was taken as \si{3\metre}. 

First, in order to study and compare the applicability of equation (\ref{eq3}) to this study’s local propagation environment on the Severn River, predictions for equation (\ref{eq3}) were scaled to a height of \si{3\metre} using equation (\ref{eq4}). Then, a comparison of the measured $C_n^2$ and the value of $C_n^2$ predicted for the same height using a combination of equation (\ref{eq3}) and equation (\ref{eq4}) is presented in \autoref{table:2}. 

\begin{table}[H]
    \centering
    \caption{
        Macro-meteorological and oceanographic parameters captured near the link (January 1\textsuperscript{st} 2020 through July 14\textsuperscript{th} 2022).
    }
    \begin{tabular}{|| m{4cm} || >{\centering}m{3cm} | >{\centering}m{3cm} | >{\centering\arraybackslash}m{3cm} ||}
        \hline
         & \textbf{Air-water Temperature Difference \si{\celsius}} & \textbf{Measured $C_n^2$} & \textbf{equation (\ref{eq3}) Predicted $C_n^2$} \\
        \hline
        \hline
        \textbf{Count} & \si{1,241,037} & \si{1,246,802} & \si{1,156,426} \\
        \textbf{Mean} & $-0.4$ & $1.845\times10^{-14}$ & $2.862\times10^{-15}$ \\
        \textbf{Std.} & $4.34$ & $3.646\times10^{-14}$ & $4.350\times10^{-15}$ \\
        \textbf{Min.} & $-12.25$ & $1.642\times10^{-17}$ & $6.200\times10^{-17}$ \\
        \textbf{\si{25\%}} & $-2.66$ & $2.871\times10^{-15}$ & $3.087\times10^{-16}$ \\
        \textbf{Median} & $-0.07$ & $8.326\times10^{-15}$ & $1.193\times10^{-15}$ \\
        \textbf{\si{75\%}} & $3.11$ & $2.091\times10^{-14}$ & $3.652\times10^{-15}$ \\
        \textbf{Max} & $18.8$ & $2.404\times10^{-12}$ & $5.339\times10^{-14}$ \\
        \hline                                                                                          
    \end{tabular}
    \label{table:2}
\end{table}

\begin{figure}[H]
    \centering
    \includegraphics[width=0.5\textwidth]{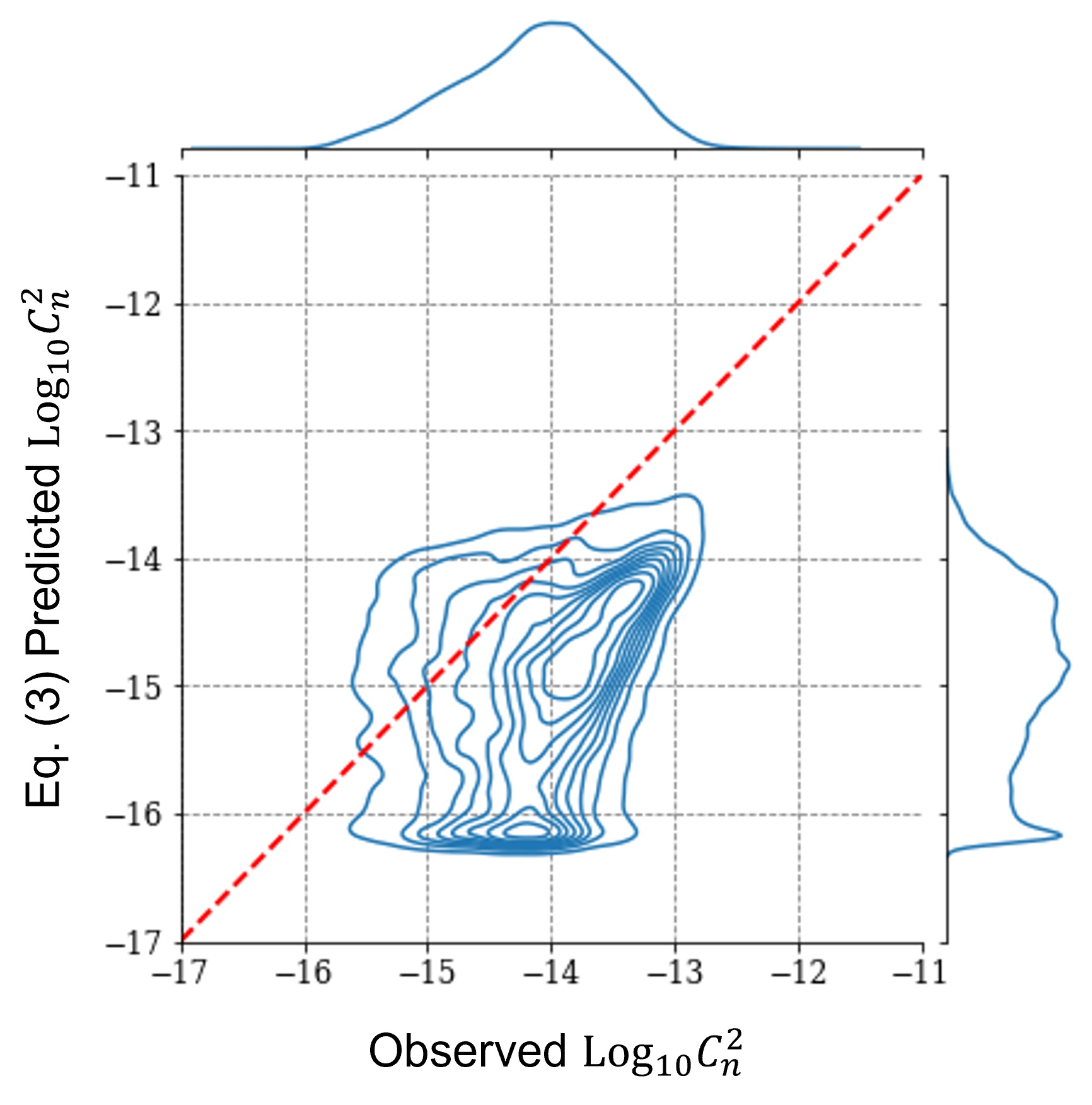}
    \caption{
        Joint distribution of measured $C_n^2$ and the macro-meteorological model in equation (\ref{eq3}) predicted $C_n^2$ at \si{3\metre} between January 1\textsuperscript{st} 2020 and July 14\textsuperscript{th} 2022.
    }
    \label{fig:2}
\end{figure}

The predicted $C_n^2$ of the model presented in \cite{chen2019climatological}, captured in \autoref{table:2} as equation (\ref{eq3}), is generally lower than the observed $C_n^2$ over the Severn River for the period between January 1\textsuperscript{st} 2020 and July 14\textsuperscript{th} 2022. This may be due to the altitude at which equation (\ref{eq3}) was developed, approximately \si{1720\metre} above sea level \cite{chen2019climatological}, or the deleterious impact of local traffic and aerosols along the Severn River propagation path. Despite this general under-prediction, equation (\ref{eq3}) provides a remarkably elegant tool for estimating local $C_n^2$ from measured air-water temperature difference alone. The prediction accuracy of equation (\ref{eq3}) is further analyzed through the joint distribution of predicted and measured $C_n^2$ in \autoref{fig:2}. 

\autoref{fig:2} highlights both the general under-prediction of $C_n^2$ in the propagation path over the Severn River, as well as the similar shape in the distributions between measured and predicted $C_n^2$. This under-prediction is most pronounced when the air-water temperature difference approaches \si{0\celsius} as identified in \cite{frederickson2006measurements}. For the period between January 1\textsuperscript{st} 2020 and July 14\textsuperscript{th} 2022, the MAE in predicted $\log_{10} C_n^2$ was $0.981$, while the mean absolute percentage error (MAPE) was \si{7.02\%}. For the period between July 14\textsuperscript{th} 2021 and July 14\textsuperscript{th} 2022, the MAE in predicted $\log_{10} C_n^2$ was $0.989$, while the MAPE was \si{7.08\%}. These metrics establish a baseline for the level of accuracy with which an observer could predict $C_n^2$ from local macro-meteorological parameters using another model, in this case equation (\ref{eq3}) which specifically utilized an air-water temperature difference, in the Severn River propagation environment. This study seeks to improve upon these predictions by developing a hybrid model, which couples both a macro-meteorological model, equation (\ref{eq3}) with a GBDT model trained on its residuals as outlined in equation (\ref{eq1}).

\subsection{Hybrid and data-only GBDT models}

\begin{figure}[H]
    \centering
    \includegraphics[width=0.8\textwidth]{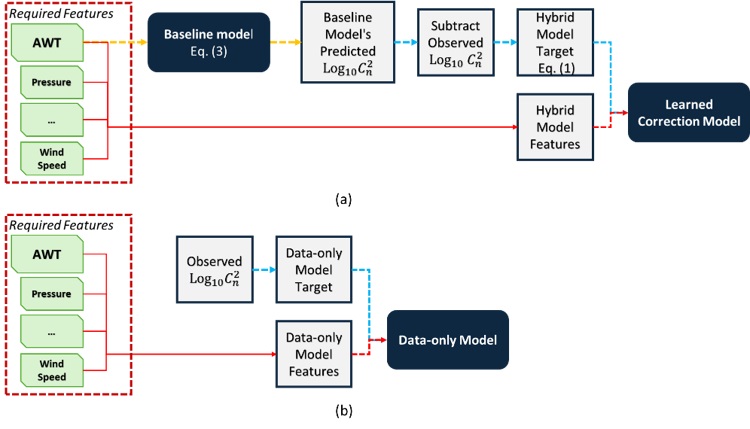}
    \caption{
        The training processes for the hybrid model (a) and the data-only model (b).
    }
    \label{fig:3}
\end{figure}

Locally measured meteorological parameters and scintillometer readings of $C_n^2$ were used to train a hybrid and a data-only model under the GBDT architecture as described in \cite{ke2017lightgbm}. Specifically, local $C_n^2$ data with a \si{1\minute} frequency was, with some infrequent dropouts, available between January 1\textsuperscript{st} 2020 and July 14\textsuperscript{th} 2022, where the period between July 14\textsuperscript{th} 2021 and July 14\textsuperscript{th} 2022 contained 460,040 observations of $C_n^2$ alongside the meteorological parameters described in \autoref{table:1} and was set aside as a long-term validation set to evaluate the baseline model in equation (\ref{eq3}), the hybrid model, and the data-only model. The remaining 696,386 observations of $C_n^2$ and local meteorological parameters were used to develop bootstrapped training samples. This serves both to estimate model performance with a given number of training observations, and to estimate the variability in validation set predictions for a model with a given number of training observations and architecture. 1440 of these \si{1\minute} observations, denoted as one “day-equivalent” of observation, were sampled randomly and with replacement from the training set, which spanned from January 1\textsuperscript{st} 2020 to July 14\textsuperscript{th} 2021. Both the hybrid and data-only models leveraged the same meteorological parameters as training features. While the data-only model was trained to predict observed $\log_{10} C_n^2$ from those features alone, the hybrid model predicted $\log_{10} C_n^2$ following equation (\ref{eq2}) from both these meteorological parameters and the baseline model in equation (\ref{eq3}), as described in \autoref{fig:3}. 

Both the hybrid and data-only models in \autoref{fig:3}. were trained to fit their target from their available features under the GBDT architecture. The GBDT architecture leverages model hyper-parameters in training to define the training and topology of constituent trees, and the way in which those trees are aggregated to form final model predictions \cite{ke2017lightgbm} \cite{pedregosa2011scikit}. An effort was made to identify reasonable hyper-parameters for a given number of days-equivalent observation using a grid search methodology \cite{pedregosa2011scikit}. The loss function was set to MAE, with 512 trees used in each ensemble, and the GBDT methodology discussed in \cite{ke2017lightgbm}. The possible hyper-parameters are described in \autoref{table:3}. 

\begin{table}[H]
    \centering
    \caption{
        Possible hyper-parameter combinations used in grid search.
    }
    \begin{tabular}{|| m{6cm} || >{\centering}m{3cm} | >{\centering\arraybackslash}m{5cm} ||}
        \hline
        \textbf{Description} & \textbf{Hyper\-parameter name \cite{ke2017lightgbm} \cite{pedregosa2011scikit}} & \textbf{Possible values} \\
        \hline
        \hline
        The number of leaves for each constituent tree node. & \texttt{num\_leaves} & 8, 128, 512, 1024, 4096 \\
        The model learning rate under the gradient boosting framework. & \texttt{learning\_rate} & 0.5, 0.2, 0.1, 0.01 \\
        The minimum number of training observations in each leaf node of each constituent tree. & \texttt{min\_data\_in\_leaf} & 32, 128, 512, 2048, 4096 \\
        \hline                                                                                          
    \end{tabular}
    \label{table:3}
\end{table}

Each possible combination of hyper-parameters in \autoref{table:3} was evaluated for every fifth number of days-equivalent observation in the training set, from 1 to 480, as described in \autoref{table:4}. The combination of hyper-parameters in \autoref{table:3} with the lowest error in training a data-only model under the GBDT architecture was selected for both the data-only and hybrid models. Selecting reasonable hyper-parameters helps to improve each model’s prediction accuracy for the given number of days-equivalent observation. The metrics, features, and target parameters are further described in \autoref{table:4}. 

\begin{table}[H]
    \centering
    \caption{
        Features and Hyper-parameters Available in Hybrid and Data-only Model Training.
    }
    \begin{tabular}{| >{\centering}m{1.5cm} | >{\centering}m{1.2cm} | >{\centering}m{1cm} | >{\centering}m{4.5cm} || >{\footnotesize}>{\centering}m{2.2cm} | >{\centering\arraybackslash}m{3.6cm} |}
        \hline
        \textbf{Model} & \textbf{Target} & \textbf{Metric} & \textbf{Features} & \textbf{Number of days-equivalent Observations in Training Set} & \textbf{Hyper-parameters }\\
        \hline
        \multirow{4}{*}{\textbf{Data-only}} & \multirow{4}{*}{\textbf{\(\log_{10} C_n^2\)}} & \multirow{4}{*}{\textbf{MAE}} & & 1 & \multirow{5}{*}{\small\begin{tabular}[c]{@{}l@{}}\texttt{learning\_rate: 0.5} \\ \texttt{min\_data\_in\_leaf: 32} \\ \texttt{num\_leaves: 8}\end{tabular}} \\
         &  &  &  & 2 &  \\
         &  &  &  & 3 &  \\
         &  &  &  & 4 &  \\
        \multirow{31}{*}{\textbf{Hybrid}} & \multirow{31}{*}{\(tc\) (\ref{eq1})} & \multirow[t]{31}{*} & \multirow[t]{35}{*}{\small\begin{tabular}[c]{@{}l@{}}Air-water Temperature Difference \\ Wind Speed \\ Humidity \\ Pressure \\ Solar Radiation \\ Temporal Hour \\ Air-water Temperature Difference\end{tabular}}  & 5 &  \\
        \cline{5-6}
         &  &  &  & 6 & \multirow{5}{*}{\small\begin{tabular}[c]{@{}l@{}}\texttt{learning\_rate: 0.1} \\ \texttt{min\_data\_in\_leaf: 128} \\ \texttt{num\_leaves: 8}\end{tabular}} \\
         &  &  &  & 7 &  \\
         &  &  &  & 8 &  \\
         &  &  &  & 9 &  \\
         &  &  &  & 10 &  \\
         \cline{5-6}
         &  &  &  & 12 & \multirow{5}{*}{\small\begin{tabular}[c]{@{}l@{}}\texttt{learning\_rate: 0.1} \\ \texttt{min\_data\_in\_leaf: 128} \\ \texttt{num\_leaves: 8}\end{tabular}} \\
         &  &  &  & 14 &  \\
         &  &  &  & 16 &  \\
         &  &  &  & 18 &  \\
         &  &  &  & 21 &  \\
         \cline{5-6}
         &  &  &  & 24 & \multirow{5}{*}{\small\begin{tabular}[c]{@{}l@{}}\texttt{learning\_rate: 0.1} \\ \texttt{min\_data\_in\_leaf: 128} \\ \texttt{num\_leaves: 8}\end{tabular}} \\
         &  &  &  & 27 &  \\
         &  &  &  & 30 &  \\
         &  &  &  & 45 &  \\
         &  &  &  & 60 &  \\
         \cline{5-6}
         &  &  &  & 75 & \multirow{5}{*}{\small\begin{tabular}[c]{@{}l@{}}\texttt{learning\_rate: 0.01} \\ \texttt{min\_data\_in\_leaf: 128} \\ \texttt{num\_leaves: 8}\end{tabular}} \\
         &  &  &  & 90 &  \\
         &  &  &  & 120 &  \\
         &  &  &  & 150 &  \\
         &  &  &  & 180 &  \\
         \cline{5-6}
         &  &  &  & 210 & \multirow{5}{*}{\small\begin{tabular}[c]{@{}l@{}}\texttt{learning\_rate: 0.01} \\ \texttt{min\_data\_in\_leaf: 512} \\ \texttt{num\_leaves: 128}\end{tabular}} \\
         &  &  &  & 240 &  \\
         &  &  &  & 270 &  \\
         &  &  &  & 300 &  \\
         &  &  &  & 330 &  \\
         \cline{5-6}
         &  &  &  & 360 & \multirow{5}{*}{\small\begin{tabular}[c]{@{}l@{}}\texttt{learning\_rate: 0.01} \\ \texttt{min\_data\_in\_leaf: 128} \\ \texttt{num\_leaves: 1024}\end{tabular}} \\
         &  &  &  & 390 &  \\
         &  &  &  & 420 &  \\
         &  &  &  & 450 &  \\
         &  &  &  & 480 & \\
         \hline
    \end{tabular}
    \label{table:4}
\end{table}

\begin{figure}[H]
    \centering
    \includegraphics[width=0.8\textwidth]{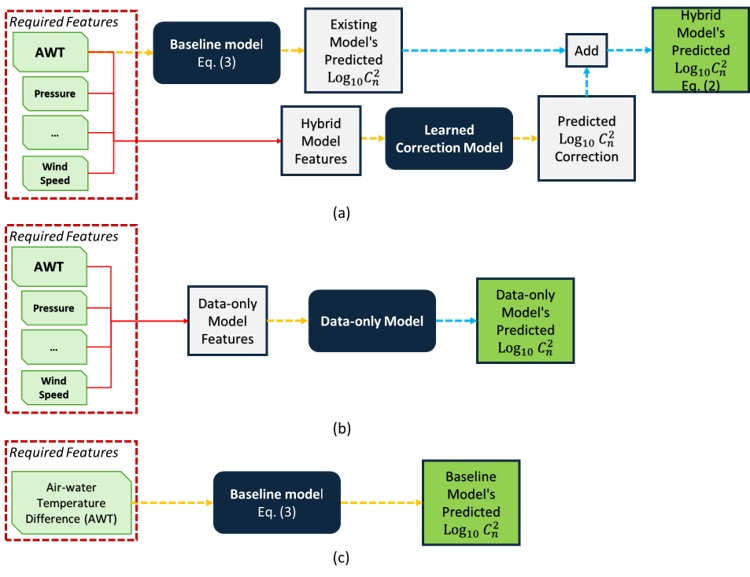}
    \caption{
        Generating predicted $\log_{10} C_n^2$ from meteorological parameters using the hybrid model (a), the data-only model (b), and the baseline model (c).
    }
    \label{fig:4}
\end{figure}

In \autoref{table:4}, both the hybrid and data-only models share the same set of features described in \autoref{table:1}, along with the temporal hour and air-water temperature difference, interpolated to a \si{1\minute} frequency. The data-only model seeks to predict $\log_{10} C_n^2$ directly from these features, where the hybrid model seeks to predict the $tc$ defined in equation (\ref{eq1}), such that it can be combined with the $\log_{10}$ of the prediction generated by the baseline macro-meteorological model in equation (\ref{eq3}) for a given observation vector to produce a hybrid $\log_{10} C_n^2$ following equation (\ref{eq2}), as demonstrated in \autoref{fig:4}.

Both hybrid and data-only models in \autoref{fig:4}. were trained and evaluated using the features, hyper-parameters, and number of days-equivalent observation in \autoref{table:4}. The hybrid and data-only models were evaluated for each number of days-equivalent observation across 20 iterations. Bootstrapped samples of the selected number of days-equivalent observation formed the training set in each iteration, and for each model. The hybrid and data-only models were evaluated under the MAE and MAPE metrics. Performance was averaged across the 20 iterations, which helped both to estimate the mean and confidence interval for the hybrid and data-only models for a given number of days-equivalent observation. 

\section{Results and Analysis}

\begin{figure}
    \centering
    \includegraphics[width=0.5\textwidth]{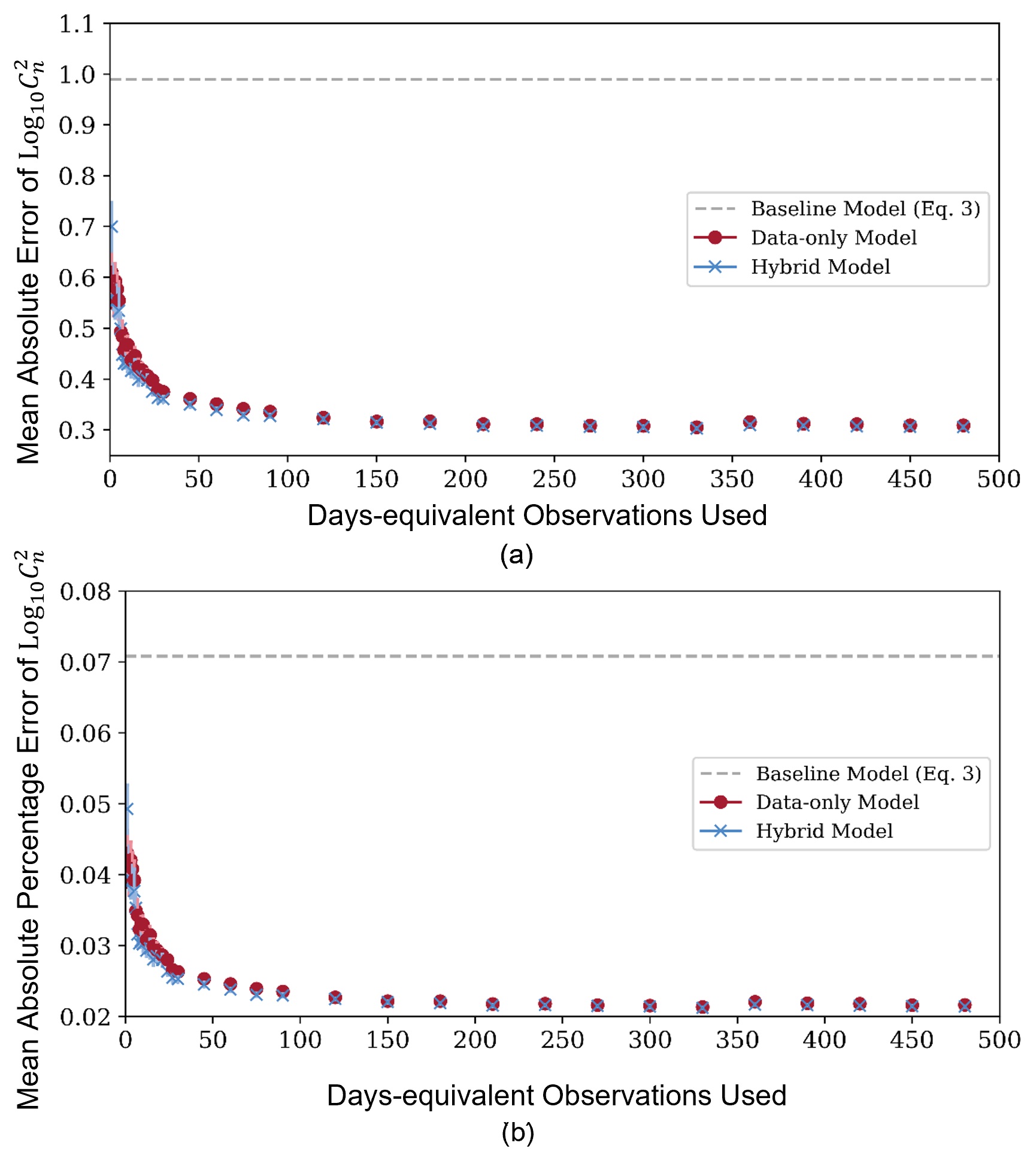}
    \caption{
        MAE (a) and MAPE (b) with \si{95\%} confidence intervals for selected numbers of bootstrapped days-equivalent observations in model training.
    }
    \label{fig:5}
\end{figure}

The bootstrapped mean and standard deviation of each metric in \autoref{table:4} was computed using the validation set, spanning one year from July 14\textsuperscript{th} 2021 through July 14\textsuperscript{th} 2022. By setting aside a one-year validation set, each model’s performance across a range of seasonal conditions was evaluated. This validation set provides an estimate of long-term prediction accuracy, for observations far removed from training data. These performance metrics are presented in \autoref{table:5} and visually in \autoref{fig:5}.

\begin{table}[H]
    \centering
    \caption{
        Model performance against the test set for hybrid and data-only models trained with select number of days-equivalent observations.
    }
    \small\begin{tabular}{| >{\centering}m{2cm} | >{\centering}m{1.2cm} | >{\centering}m{1.2cm}| >{\centering}m{1.2cm}| >{\centering}m{1.2cm}| >{\centering}m{1.2cm}| >{\centering}m{1.2cm}| >{\centering}m{1.2cm}|>{\centering\arraybackslash}m{1.2cm} |}
        \hline              
        \multirow{3}{2cm}{\small Number of days-equivalent Observations in Training Set} & \multicolumn{4}{c|}{Mean Absolute \si{\%} Error (validation set)} & \multicolumn{4}{c|}{MAE $\log_{10} C_n^2$ (validation set)} \\
        \cline{2-9}
         & \multicolumn{2}{c|}{Hybrid Model} & \multicolumn{2}{c|}{Data-only Model} & \multicolumn{2}{c|}{Hybrid Model} & \multicolumn{2}{c|}{Data-only Model} \\
         \cline{2-9}
         & Bootstrap mean & Bootstrap Std. & Bootstrap mean & Bootstrap Std. & Bootstrap mean & Bootstrap Std. & Bootstrap mean & Bootstrap Std. \\
        \hline 
        1 & 4.92\% & 0.78\% & 4.30\% & 0.74\% & 0.700 & 0.110 & 0.610 & 0.109 \\
        2 & 4.10\% & 0.70\% & 3.89\% & 0.43\% & 0.581 & 0.102 & 0.549 & 0.057 \\
        3 & 3.91\% & 0.54\% & 4.21\% & 0.58\% & 0.553 & 0.076 & 0.593 & 0.080 \\
        4 & 3.91\% & 0.50\% & 4.08\% & 0.52\% & 0.555 & 0.071 & 0.577 & 0.073 \\
        5 & 3.77\% & 0.71\% & 3.92\% & 0.61\% & 0.534 & 0.101 & 0.555 & 0.085 \\
        6 & 3.53\% & 0.52\% & 3.49\% & 0.39\% & 0.499 & 0.073 & 0.493 & 0.053 \\
        7 & 3.16\% & 0.44\% & 3.43\% & 0.50\% & 0.448 & 0.061 & 0.484 & 0.069 \\
        8 & 3.03\% & 0.14\% & 3.23\% & 0.25\% & 0.430 & 0.018 & 0.457 & 0.033 \\
        9 & 3.07\% & 0.23\% & 3.31\% & 0.24\% & 0.435 & 0.032 & 0.468 & 0.033 \\
        10 & 3.02\% & 0.24\% & 3.30\% & 0.27\% & 0.429 & 0.033 & 0.468 & 0.037 \\
        12 & 2.93\% & 0.19\% & 3.08\% & 0.28\% & 0.416 & 0.027 & 0.437 & 0.038 \\
        14 & 2.96\% & 0.30\% & 3.15\% & 0.29\% & 0.419 & 0.042 & 0.445 & 0.039 \\
        16 & 2.80\% & 0.18\% & 2.99\% & 0.25\% & 0.398 & 0.025 & 0.424 & 0.034 \\
        18 & 2.85\% & 0.16\% & 2.94\% & 0.24\% & 0.405 & 0.021 & 0.417 & 0.033 \\
        21 & 2.80\% & 0.20\% & 2.87\% & 0.20\% & 0.397 & 0.028 & 0.406 & 0.027 \\
        24 & 2.63\% & 0.11\% & 2.80\% & 0.12\% & 0.375 & 0.016 & 0.398 & 0.016 \\
        27 & 2.55\% & 0.14\% & 2.66\% & 0.14\% & 0.363 & 0.020 & 0.379 & 0.019 \\
        30 & 2.53\% & 0.09\% & 2.63\% & 0.11\% & 0.360 & 0.013 & 0.375 & 0.015 \\
        45 & 2.45\% & 0.08\% & 2.53\% & 0.09\% & 0.349 & 0.011 & 0.361 & 0.012 \\
        60 & 2.38\% & 0.05\% & 2.46\% & 0.06\% & 0.339 & 0.008 & 0.350 & 0.009 \\
        75 & 2.31\% & 0.05\% & 2.39\% & 0.07\% & 0.329 & 0.008 & 0.341 & 0.009 \\
        90 & 2.30\% & 0.05\% & 2.35\% & 0.05\% & 0.327 & 0.008 & 0.336 & 0.007 \\
        120 & 2.25\% & 0.03\% & 2.27\% & 0.04\% & 0.321 & 0.005 & 0.324 & 0.006 \\
        150 & 2.20\% & 0.04\% & 2.22\% & 0.05\% & 0.314 & 0.006 & 0.316 & 0.008 \\
        180 & 2.19\% & 0.03\% & 2.22\% & 0.03\% & 0.312 & 0.005 & 0.317 & 0.005 \\
        210 & 2.15\% & 0.03\% & 2.18\% & 0.03\% & 0.307 & 0.004 & 0.311 & 0.004 \\
        240 & 2.16\% & 0.03\% & 2.18\% & 0.04\% & 0.308 & 0.004 & 0.311 & 0.006 \\
        270 & 2.15\% & 0.03\% & 2.16\% & 0.04\% & 0.306 & 0.004 & 0.309 & 0.006 \\
        300 & 2.14\% & 0.03\% & 2.16\% & 0.04\% & 0.306 & 0.004 & 0.308 & 0.005 \\
        330 & 2.12\% & 0.03\% & 2.14\% & 0.03\% & 0.303 & 0.004 & 0.305 & 0.004 \\
        360 & 2.17\% & 0.02\% & 2.21\% & 0.03\% & 0.309 & 0.003 & 0.315 & 0.004 \\
        390 & 2.16\% & 0.02\% & 2.19\% & 0.02\% & 0.308 & 0.003 & 0.312 & 0.004 \\
        420 & 2.15\% & 0.01\% & 2.18\% & 0.03\% & 0.307 & 0.002 & 0.311 & 0.004 \\
        450 & 2.15\% & 0.02\% & 2.16\% & 0.02\% & 0.306 & 0.002 & 0.309 & 0.003 \\
        480 & 2.14\% & 0.02\% & 2.16\% & 0.02\% & 0.306 & 0.003 & 0.309 & 0.004 \\
        \hline
    \end{tabular}
    \label{table:5}
\end{table}

The equation (\ref{eq3}) baseline model’s predictions over the validation set were presented alongside \autoref{fig:5}, with the MAE in predicted $\log_{10} C_n^2$ calculated as $0.989$ and the MAPE was $7.08\%$. Based on the results in \autoref{table:5}, the existing model can be improved using the hybrid model framework using only one day’s worth of data, or a block of 1440 sequential \si{1\minute} observations. The MAE in predicted $\log_{10} C_n^2$ trained on one days-equivalent observation for the hybrid model was estimated at $0.700$, a \si{29\%} reduction compared against the existing model. Similarly, the data-only model’s reduction in error was estimated at $0.611$, a \si{38\%} reduction compared against the existing model. 

\begin{figure}[H]
    \centering
    \includegraphics[width=0.5\textwidth]{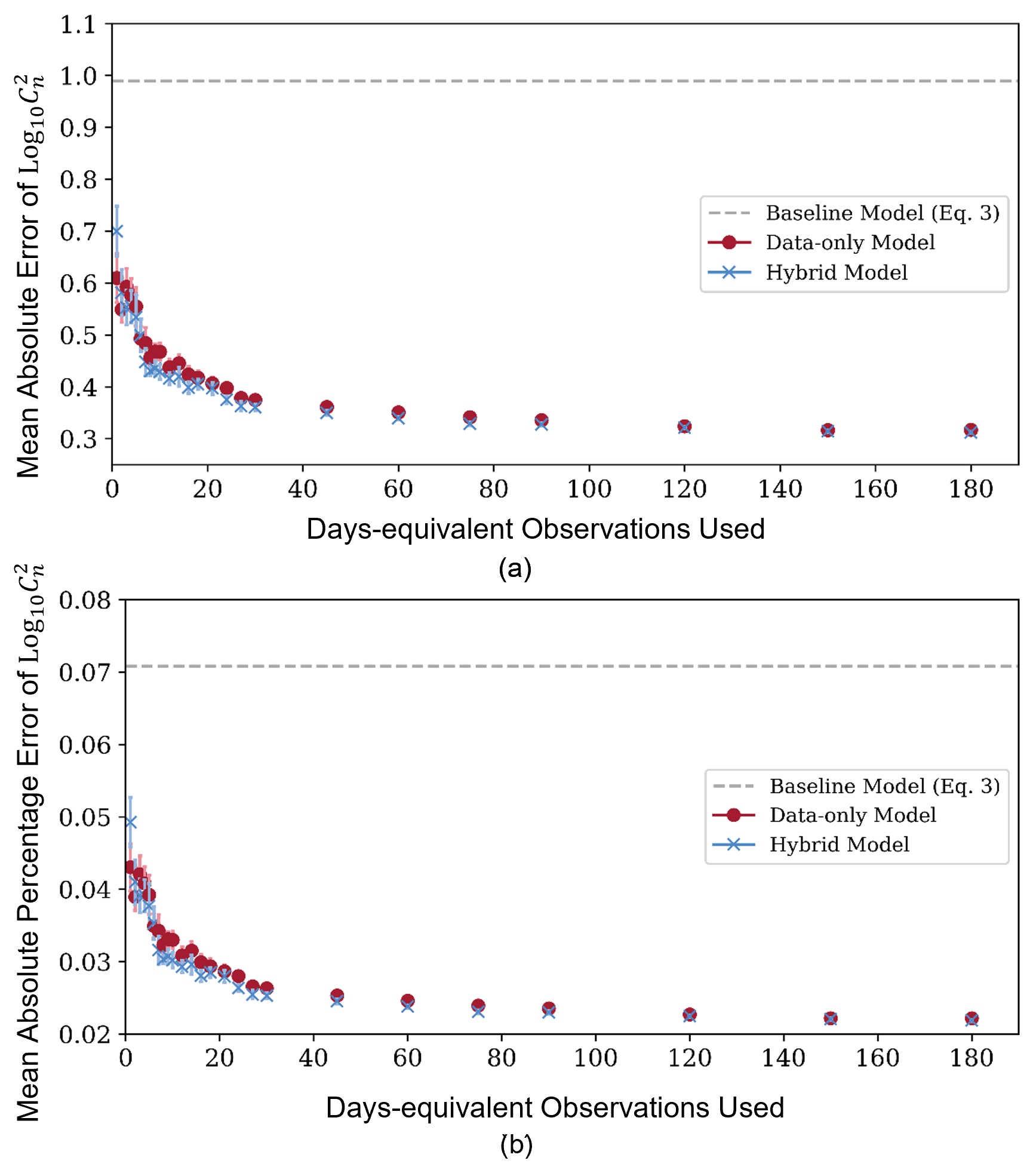}
    \caption{
        MAE (a) and MAPE (b) with \si{95\%} confidence intervals for hybrid and data-only models trained with up to 180 bootstrap-sampled days-equivalent data.
    }
    \label{fig:6}
\end{figure}

Both the hybrid and data-only models show similar gains in prediction accuracy when compared against the existing macro-meteorological model. These gains appear to level off somewhat after 180 sampled days-equivalent of observation. With 180 sampled days-equivalent of observation, the hybrid model had an estimated \si{68\%} reduction in error, improving to a \si{69\%} reduction in error with 480 sampled days-equivalent of observation. It is interesting to note that the hybrid model appears to minimally outperform the data-only model over much of the investigation period, but when the days-equivalent observation is less than 5, the data-only model marginally outperforms the hybrid model. From that point, the hybrid model is marginally more effective, with gains in performance falling below \si{2\%} near the 24 days-equivalent observation mark. The models trained with 1 to 180 days-equivalent, are presented in greater detail in \autoref{fig:6}. 

\begin{figure}[H]
    \centering
    \includegraphics[width=0.7\textwidth]{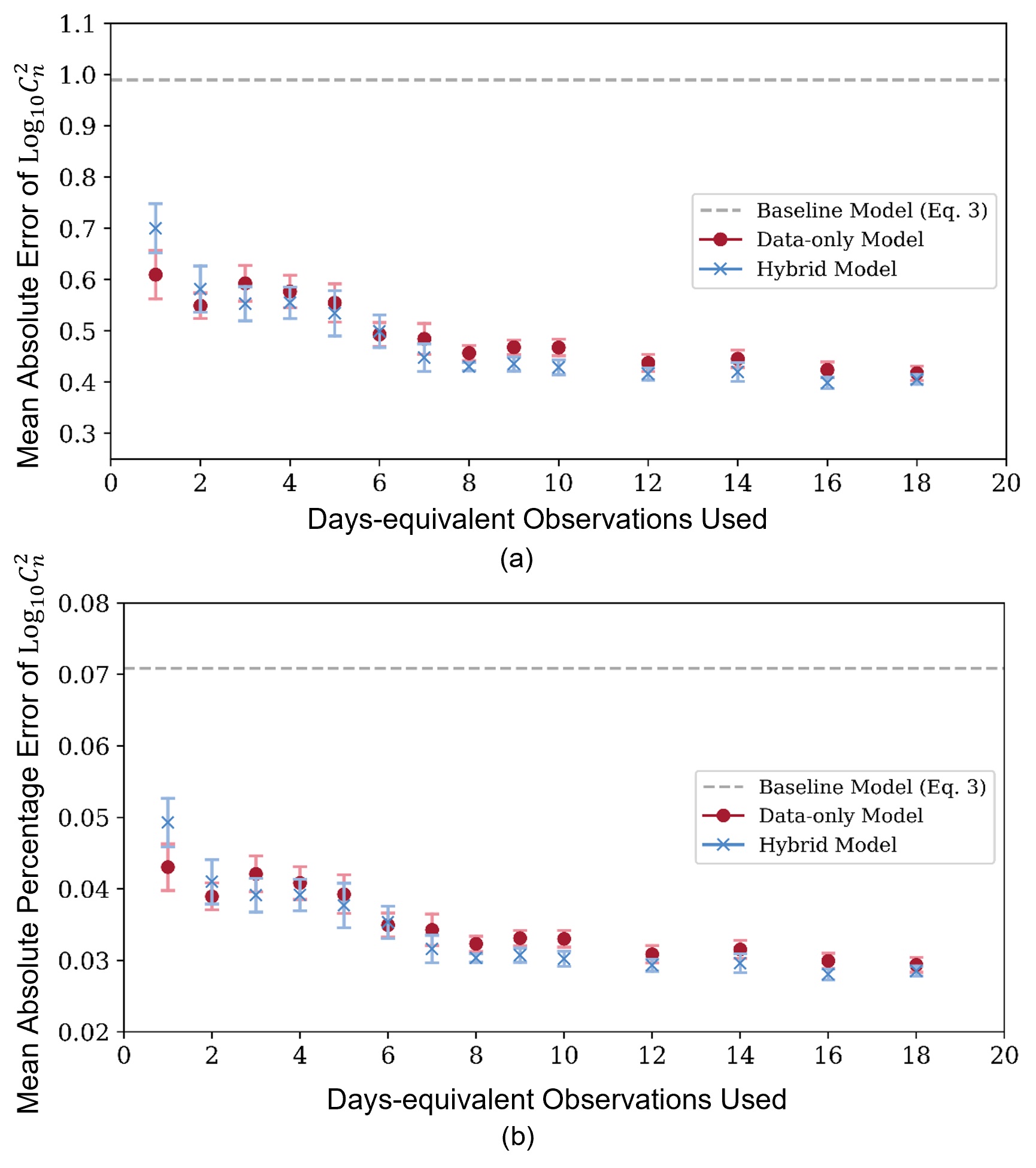}
    \caption{4
        MAE (a) and MAPE (b) with \si{95\%} confidence intervals for hybrid and data-only models trained with up to 18 days of bootstrap-sampled days-equivalent data.
    }
    \label{fig:7}
\end{figure}

\autoref{fig:5} and \autoref{fig:6} highlight three key results from this investigation. The first is that hybrid models, as well as a data-only model, can significantly improve on a selected baseline model using as little as one day’s observation from the local microclimate. The performance of the hybrid model was often within the confidence interval of the data-only model trained on the same number of bootstrapped samples. This convergence in performance is most evident in models trained with more than 18 days-equivalent of bootstrapped observations, as seen in \autoref{fig:7}. 

In \autoref{fig:7}, performance relative to equation (\ref{eq3}) improves by approximately \si{58\%} under the data-only model and \si{59\%} under the hybrid model after 18 days-equivalent observation. \autoref{fig:5}, \autoref{fig:6} and \autoref{fig:7} highlight that both the data-only and hybrid models demonstrated improvements in prediction accuracy as more data was made available for training. However, the rate of improvement plateaus as more training observations are made available. This effect only appeared to slow substantially after at least 180 days-equivalent of bootstrapped samples were used in fitting the models. 

\begin{figure}[H]
    \centering
    \includegraphics[width=0.9\textwidth]{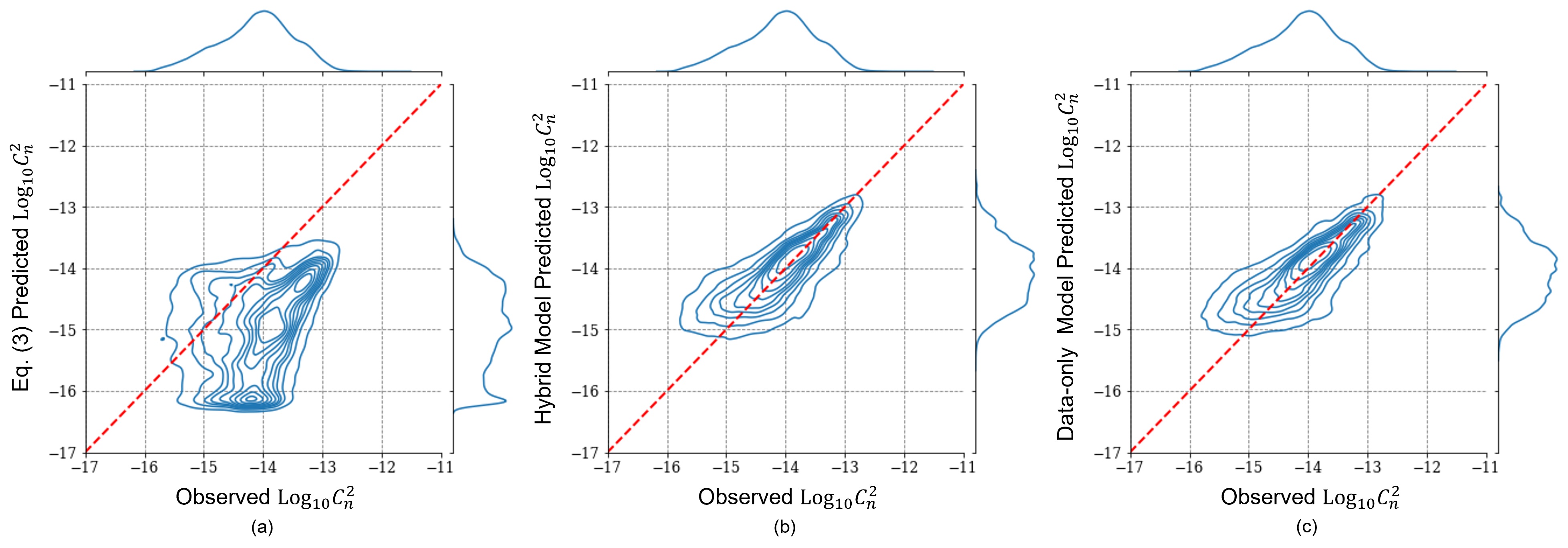}
    \caption{
        Joint distribution of measured $C_n^2$ and the baseline model’s predicted $C_n^2$ from equation (\ref{eq3}) at \si{3\metre} on the validation set in (a) with hybrid model predictions in (b), and with a data-only model in (c).
    }
    \label{fig:8}
\end{figure}

Both the hybrid and data-only models are capable of leveraging in-situ measurements of turbulent effects and meteorological parameters in generating improved predictions of $C_n^2$ when compared to the baseline model. With only one days-equivalent of observation, this performance improvement is estimated at approximately \si{29\%} in MAE and MAPE of $\log_{10} C_n^2$. As more data is made available for training the data-only and hybrid models, this effect becomes more pronounced, with prediction error falling by approximately \si{68\%} when 180 days-equivalent or more is available. To better understand the performance of the hybrid and data-only models, and their performance relative to the baseline model in equation (\ref{eq3}), one of each model was trained using all available measurements in the training set. The hyper-parameters in the last row of \autoref{table:4} were selected for training. The improvement in prediction accuracy for validation set observations is captured in \autoref{fig:8}. 

In \autoref{fig:8}, the hybrid model serves to adjust predictions for $C_n^2$ based on locally measured macro-meteorological parameters. When aggregated across the validation set, the hybrid model’s prediction distribution in \autoref{fig:8} (b) more closely matches the observed distribution of $\log_{10} C_n^2$ than the initial prediction distribution in \autoref{fig:8} (a). While it does not fully capture the relationships between the local propagation environment and observed $C_n^2$ , the hybrid model presents an improvement in aggregate prediction accuracy over the baseline in equation (\ref{eq3}) when evaluated over the one-year validation set. 

In addition to developing a hybrid model from the baseline model in equation (\ref{eq3}), two additional literature models were augmented under the hybrid model framework to investigate the framework’s extensibility. The macro-meteorological model presented in \cite{sadot1992forecasting} was trained for an over-land propagation environment at a height of \si{15\metre}, and captures diurnal variation in $C_n^2$. In order to generate predictions from local meteorological data, the dynamic range presented in \cite{sadot1992forecasting} was applied, with measurements outside of that dynamic range dropped from the training and validation sets. This model is presented in parametric form as:

\begin{equation}
    \begin{aligned}
        C_n^2 & = (3.8 \times 10^{-14})W + f(T) + f(U) + f(RH) ― (5.3 \times 10^{-13}) \\
        \text{where} \\
        f(T) & = (2.0 \times 10^{-15})T \\
        f(U) & = ( -2.5 \times 10^{-15})U + (1.2 \times 10^{-15})U2 ― (8.5 \times 10^{-15})U3 \\
        f(RH) & = ( -2.8 \times 10^{-15})RH + (2.9 \times 10^{-17})RH2 ― (1.1 \times 10^{-19})RH3 \\
    \end{aligned}
\end{equation}
\label{eq5}

In equation (\ref{eq5}), $W$ denotes the temporal hour weight \cite{sadot1992forecasting}, $T$  denotes the temperature in \si{\kelvin}, $RH$ denotes the relative humidity in \si{\%}, and $U$ denotes the wind speed in \si{\metre\per\second}. Over the validation set between July 2021 and July 2022, equation (\ref{eq5}) had a MAE in predicting $\log_{10} C_n^2$ of $1.068$. Using equation (\ref{eq5}) as a baseline model, a hybrid model was developed to augment the predictions of \ref{eq5} to the Severn River’s microclimate. The scaling law in equation (\ref{eq4}) was applied to generate predictions for a height of \si{3\metre}. With one days-equivalent observation, the MAE on the one-year validation set as reduced by \si{38\%}. After 7 days-equivalent observation, the improvement was \si{55\%}, growing to \si{69\%} after 180 days-equivalent observation.  The effectiveness of the hybrid model approach in this context may be due to the greater disparity in the environment for which equation (\ref{eq5}) was developed, with a focus on over-land propagation rather than over-water, as in equation (\ref{eq3}) \cite{sadot1992forecasting} \cite{chen2019climatological}. 

The model described in equation (\ref{eq5}) was analyzed and refit in \cite{wang2015prediction} to better capture turbulent dynamics in a coastal environment. The “Offshore macrometeorological model of $C_n^2$” described in \cite{wang2015prediction} is reproduced as:
\begin{equation}
    \begin{aligned}
        C_n^2 & = (―1.58 \times 10^{-15})W + f(T) + f(U) + f(RH) ― (7.44 \times 10^{-14}) \\
        \text{where} \\
        f(T) & = (2.74 \times 10^{-16})T \\
        f(U) & = (3.37 \times 10^{-16})U + (1.92 \times 10^{-16})U^2 ― (2.8 \times 10^{-17})U^3 \\
        f(RH) & = (8.3 \times 10^{-17})RH - (2.22 \times 10^{-18})RH^2 + (1.42 \times 10^{-20})RH^3 \\
    \end{aligned}
\end{equation}
\label{eq6}

In equation (\ref{eq6}), $W$ denotes the temporal hour weight \cite{sadot1992forecasting} $T$ denotes the temperature in \si{\kelvin}, $RH$ denotes the relative humidity in \si{\%}, and $U$ denotes the wind speed in \si{\metre\per\second}.  As for equation (\ref{eq5}), measurements outside the dynamic range presented in \cite{wang2015prediction} were removed from the training and validation sets. This model demonstrated lower prediction error than that in equation (\ref{eq5}) over the one-year validation set, with a MAE in predicted $\log_{10} C_n^2$ of $0.533$.  Taking equation (\ref{eq6}) as the baseline, the hybrid model approach failed to improve prediction accuracy in the one-days equivalent observation case but improved MAE by \si{8\%} at 7 days-equivalent observation and by \si{40\%} at 180 days-equivalent observation. The hybrid model approach with equation (\ref{eq6}) as a baseline could indicate that, for baseline models well suited to the environments in which they are applied, more local measurement is required to improve prediction accuracy.

\section{Conclusions} 

Macro-meteorological models which generate predicted $C_n^2$ from locally measured parameters present a useful baseline for efficiently estimating local turbulent effects. These macro-meteorological models often fail to capture the full extent of turbulent dynamics when applied in new propagation environments, such as the air-water boundary layer above the Severn River. These challenges motivated the development of the hybrid model framework for augmenting baseline model predictions with corrections learned from a minimal amount of local observation. This hybrid model framework approach is investigated in detail with one selected baseline model, and then evaluated with two additional baseline macro-meteorological models over a single propagation path. The hybrid model framework approach itself is not specific to any baseline macro-meteorological model, architecture, or microclimate, and it may demonstrate similar performance improvements when extended to new baseline models, architectures, and domains. Both the hybrid model and the data-only model outperformed the baseline model, in some cases when only one day-equivalent of \si{1\minute} observations was available for training. 

The hybrid model framework effectively augmented three baseline models, improving their prediction accuracy over a one-year validation set. For the equation (\ref{eq3}) baseline, both the hybrid and data-only model’s demonstrated similar performance and predictive power for a given number of bootstrapped samples, with the hybrid model marginally outperforming the data-only model after approximately the 5 days-equivalent observation mark. The hybrid model’s improved steadily through approximately 180 days-equivalent observation, and marginally thereafter. With only one days-equivalent of observation, the hybrid model’s performance improvement is an estimated reduction in MAE of approximately \si{29\%}, which grows to approximately \si{68\%} with 180 days-equivalent of observation. The absence of a total performance asymptote is potentially indicative of the seasonal variation in the local micro-climate and its impact on $C_n^2$ over the propagation path. 

While these models showed a remarkable increase in prediction accuracy when compared to the baseline models, architectures which better leverage temporal dependencies, the sequential nature of the data, or better handle missing values may provide a source of further improvement. Further, as highlighted by the amount of data required to observe a possible asymptote in performance improvement, the seasonality of the local propagation environment merits further study. Its impact on the development of new models, especially when data is limited, may help explain the relationship between validation set prediction performance and the number of bootstrapped samples used in training the models.

\section*{Funding} 

This work is supported in part by the Office of Naval Research, the Directed Energy Joint Technology Office, and the United States Naval Academy Trident Scholar Program.

\section*{Acknowledgments} 

The authors would also like to thank the meteorologists at the National Data Buoy Center for making their data available, and the team at the Water Front Readiness Center in Annapolis, MD for their support in establishing the scintillometer link.

\section*{Data availability} 

Data underlying the results presented in this paper are not publicly available at this time but may be obtained from the authors upon reasonable request.


\bibliographystyle{unsrtnat}   

\bibliography{2_references}   

\end{document}